# MAGNET DESIGNS FOR MUON COLLIDER RING AND INTERACTION REGIONS*

A.V. Zlobin[#], Y.I. Alexahin, V.V. Kashikhin, N.V. Mokhov, Fermilab, Batavia, IL 60510, U.S.A.

*Abstract*

Conceptual designs of superconducting magnets for the storage ring of a Muon Collider with a 1.5 TeV c.o.m. energy and an average luminosity of $10^{34}$ cm$^{-2}$s$^{-1}$ are presented. All magnets are based on Nb$_3$Sn superconductor and designed to provide an adequate operating field/field gradient in the aperture with the critical current margin required for reliable magnet operation in the machine. Magnet cross-sections were optimized to achieve the accelerator field quality in the magnet aperture occupied with beams. The magnets and corresponding protective measures are designed to handle about 0.5 kW/m of dynamic heat load from the muon beam decays. Magnet parameters are reported and compared with the requirements.

## INTRODUCTION

A Muon Collider is seen as a promising machine for the future of high energy physics [1]. Particle collisions in the Muon Collider will occur through the intersection of two circulating beams inside a storage ring. Requirements and operating conditions for a Muon Collider pose significant challenges to superconducting magnet designs and technologies [2]. For instance, contrary to proton machines, the ring dipole magnets should allow the muon decay products to escape the magnet helium volume. The IR quadrupoles must have a large aperture to accommodate the large decay deposition and large beam size at the expected $\beta^*$. For the IR dipoles, the required good field quality region needs to have a vertical aspect ratio of 2:1. This imposes additional challenges for the magnet design.

This paper summarizes the results of conceptual design studies of superconducting magnets for the storage ring of a Muon Collider with a 1.5 TeV c.o.m. energy and an average luminosity of $10^{34}$ cm$^{-2}$s$^{-1}$. These studies included the choice of superconductor and magnet designs to achieve the required field or field gradient in MC Storage Ring magnets (dipoles and quadrupoles) within the specified apertures with appropriate operating margins and accelerator field quality.

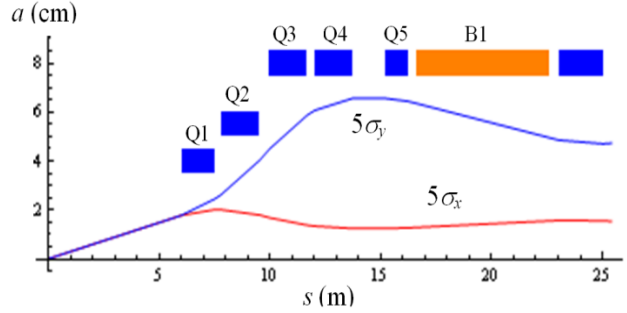

Figure 1: MC IR layout and beam size in magnets.

## MAGNET REQUIREMENTS

Muon Collider target parameters are summarized in Table 1. The storage ring lattice and the IR layout consistent with these parameters were developed and reported in [3].

The MC storage ring is based on 10 T dipole magnets. The small transverse beam size ($\sigma$~0.5 mm) requires a small aperture only ~10 mm in diameter. However, the muon decay particles and the 0.5 kW/m dynamic heat load associated with them and localized in the horizontal direction on the inner side of the storage ring, need to be intercepted outside of the magnet helium vessel on a safe distance from primary beams.

The final focus of muon beams is provided by quadrupole doublets formed by five short quadrupole magnets Q1-Q5. The bending dipoles B1, placed immediately after the final-focus doublet, generate a large dispersion function at the location of the sextupole nearest to the IP to compensate for the vertical chromaticity. The IR layout with the vertical and horizontal beam size variations is shown in Fig. 1.

The IR quadrupoles are divided into short pieces Q1-Q4 to provide space for protecting tungsten masks. The space between the Q4 and Q5 is reserved for beam diagnostics and correctors. The IR magnet design parameters are summarized in Table 2. The aperture of the magnets is determined by the following criterion $D_{x/y} = 10\sigma_{max} + 20$ mm.

Table 1: MC Storage Ring Parameters.

| Parameter | Unit | Value |
|---|---|---|
| Beam energy | TeV | 0.75 |
| Nominal dipole field | T | 10 |
| Circumference | km | 2.5 |
| Momentum acceptance | % | ±1.2 |
| Transverse emittance, $\varepsilon_N$ | $\pi$·mm·mrad | 25 |
| Number of IPs | | 2 |
| $\beta^*$ | cm | 1 |

Table 2: IR Magnet Parameters.

| Magnet type | Magnetic length, m | Magnet aperture $D_{x/y}$, mm | $G_{op}$ ($B_{op}$), T/m (T) |
|---|---|---|---|
| Q1 | 1.5 | 70 | 250 |
| Q2 | 1.7 | 100 | 187 |
| Q3 | 1.7 | 150 | 130 |
| Q4 | 1.7 | 150 | 130 |
| Q5 | 1.0 | 150 | 95 |
| B1 | 6.0 | 75/150 | (8) |

___________________________________________

* Work supported by Fermi Research Alliance, LLC under Contract DE-AC02-07CH11359 with the U.S. DOE.

[#] zlobin@fnal.gov

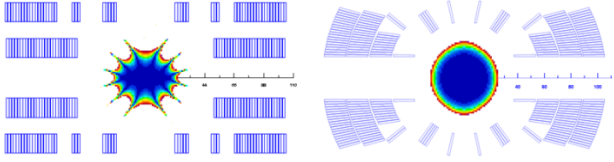

Figure 2: MC Storage Ring dipole based on 4-layer block-type coil (left) or 4-layer shell-type coil (right).

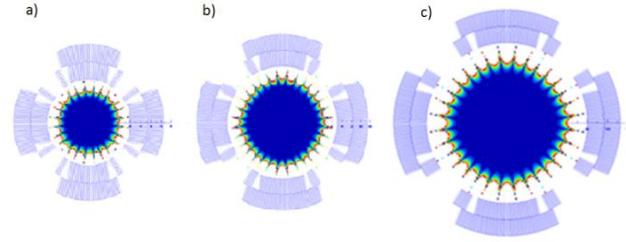

Figure 3: Q1 (a), Q2 (b), Q3-Q5 (c) cross-sections.

## MAGNET DESIGNS AND PARAMETERS

The level of magnetic fields in storage ring magnets suggests using $Nb_3Sn$ superconductor, which has the most appropriate combination of the critical parameters $J_c$, $T_c$, $B_{c2}$ and is commercially produced at the present time in long length. $Nb_3Sn$ strand and cable parameters used in this study are summarized in Table 3.

Table 3: Cable Parameters.

| Parameter | Unit | Cable 1 | Cable 2 |
|---|---|---|---|
| Number of strands | | 37 | 28 |
| Strand diameter | mm | 0.80 | 1.00 |
| Cable inner thickness | mm | 1.63 | 1.80 |
| Cable outer thickness | mm | 1.84 | 1.80 |
| Cable width | mm | 16.32 | 14.70 |
| Cu/nonCu ratio | | 1.17 | 1.00 |
| Jc(12T, 4.2K) | A/mm$^2$ | 2750 | 2750 |

### Storage Ring Dipole

The dipole requirements and operating conditions call for an open mid-plane design approach [4] and a beam pipe with large horizontal size. Cross-sections of MC Storage Ring dipoles based on 4-layer block-type (left) and shell-type (right) coils are shown in Fig.2. The main magnet parameters are summarized in Table 4.

The shell-type coil is based on keystoned Cable 1 and an iron yoke with an inner diameter (ID) of 240 mm. The block-type coil uses rectangular Cable 2 and an yoke with an ID of 250 mm. The mid-plane coil-to-coil gap in both designs is 30 mm to provide a mid-plane open space of at least 10 mm. Both designs have practically the same conductor volume and provide the maximum field in an aperture of ~11.2 T which corresponds to ~11% margin with respect to the nominal operating field at 4.5 K.

Presented dipole magnet designs have quite large horizontal and vertical Lorentz force components which lead to a high stress level in the coil. Both components need to be supported by an adequate mechanical support structure to minimize turn motion which may cause magnet quench and field quality degradation.

Table 4: Storage Ring Dipole Parameters.

| Parameter | Block design | Shell design |
|---|---|---|
| $B_{max}$ coil at 4.5K (T) | 13.37 | 13.13 |
| $B_{max}$ at 4.5 K (T) | 11.24 | 11.24 |
| $B_{op}$ (T) | 10.0 | 10.0 |
| Inductance at $B_{op}$ (mH/m) | 6.72 | 9.52 |
| Stored energy at $B_{op}$ (kJ/m) | 1280 | 1100 |
| $F_x$ at $B_{op}$ (kN/m) | 4084 | 3990 |
| $F_y$ at $B_{op}$ (kN/m) | -2216 | -1870 |

Geometrical field harmonics for both designs are reported in Table 5. In the shell-type design the accelerator field quality is achieved within a 50 mm circle and in the block-type design within an ellipse with 40 mm horizontal and 20 mm vertical size (blue area in Fig.2). In both cases it is sufficient to accommodate the orbit sagitta and large dispersion contribution to the beam size.

Table 5: Geometrical Harmonics at $R_{ref}$=10 mm ($10^{-4}$).

| Harmonic # | Block design | Shell design |
|---|---|---|
| $b_3$ | -0.081 | 0.001 |
| $b_5$ | 0.091 | -0.001 |
| $b_7$ | -0.013 | 0.002 |
| $b_9$ | -0.002 | -0.005 |

### Large-aperture IR Quadrupoles

Based on Table 2 the IR doublet needs quadrupoles with three different apertures and nominal gradients corresponding to Q1, Q2 and Q3-Q4. Q5 will operate at a lower gradient or at the same gradient but be proportionally shorter. Three basic IR quadrupole cross-sections are shown in Fig. 3, and magnet parameters are summarized in Table 6. The quadrupole apertures shown in Table 6 were increased with respect to those in Table 2 by an additional 10 mm to provide adequate space for the beam pipe, annular helium channel, optional inner absorber (liner) and possibility to shift the quadrupole axis to create a vertical bending field in quadrupoles [5].

The IR quadrupoles are based on 2-layer shell-type coils and a cold iron yoke separated from the coils by a 10 mm spacer. All the designs use Cable 1 (see Table 3). As can be seen, all the magnets provide ~12% operating margin at 4.5 K. If necessary, the margin can be increased by adding additional coil layers or operating the IR quadrupoles at 1.9 K.

Geometrical field harmonics for IR quadrupoles Q1-Q5 are presented in Table 7. The accelerator field quality is achieved within the circles (blue areas in Fig. 3) equal to 2/3 of the corresponding coil aperture.

Table 6: IR Quadrupole Parameters.

| Parameter | Q1 | Q2 | Q3-Q5 |
|---|---|---|---|
| Aperture (mm) | 80 | 110 | 160 |
| $B_{max}$ coil at 4.5 K (T) | 12.76 | 13.19 | 13.49 |
| $G_{max}$ apert at 4.5 K (T/m) | 281.5 | 209.0 | 146.0 |
| $G_{op}$ (T/m) | 250 | 187 | 130 |
| Inductance at $G_{op}$ (mH/m) | 3.57 | 6.58 | 12.88 |
| Stored energy at $G_{op}$ (kJ/m) | 493.0 | 771.3 | 1391.8 |
| $F_x$ at $G_{op}$ (kN/m) | 1790 | 2225 | 2790 |
| $F_y$ at $G_{op}$ (kN/m) | -2180 | -2713 | -3380 |

Table 7: Geometrical Harmonics at $R_{ref}$ ($10^{-4}$).

| Harmonic # | Q1 | Q2 | Q3-Q5 |
|---|---|---|---|
| $R_{ref}$ (mm) | 27 | 37 | 53 |
| $b_6$ | 0.000 | 0.000 | 0.000 |
| $b_{10}$ | -0.034 | 0.002 | 0.002 |
| $b_{14}$ | 0.862 | 0.090 | 0.086 |

### IR Dipole

The large vertical beam size in the IR region (see Fig.1) makes the parameters of the IR dipole B1 very challenging. As for the Storage ring dipole, it is important for the IR dipoles to have an open mid-plane to avoid showering of muon decay electrons in the vicinity of the superconducting coils as well as to reduce background fluxes in a detector central tracker. To remove 95% of the radiation from the aperture to the external absorber the open mid-plane gap in B1 should be at least $5\sigma_y$. The large 160 mm aperture and the large 6 cm gap limit the magnet nominal field which can be achieved with $Nb_3Sn$ coils and make it difficult to achieve an acceptable field quality in the area occupied by beams.

The cross-sections of an IR dipole based on 4-layer shell-type coil and an iron yoke with the ID of 320 mm is shown in Fig.4. The dipole parameters are summarized in Table 8.

Table 8: IR Dipole Parameters.

| Parameter | Value |
|---|---|
| Aperture (mm) | 160 |
| $B_{max}$ in coil at 4.5 K (T) | 13.03 |
| $B_{max}$ in aperture at 4.5 K (T) | 9.82 |
| $B_{op}$ (T) | 8.0 |
| Inductance at $B_{op}$ (mH/m) | 15.89 |
| Stored energy at $B_{op}$ (kJ/m) | 1558 |
| $F_x$ at $B_{op}$ (kN/m) | 3960 |
| $F_y$ at $B_{op}$ (kN/m) | -1650 |

The dipole design is based on Cable 1 and provides the maximum design field in an aperture of 9.82 T at 4.5 K, which corresponds to ~23% margin with respect to the nominal field of 8 T. Note that the maximum field in the coil is as high as 13 T. The shell-type coil design was chosen due to its better ratio between the magnet aperture and the mid-plane gap. Studies of alternative magnet design approaches for B1 will continue. The IR dipole has a high level of Lorentz forces as the storage ring dipole.

Geometrical field harmonics for IR dipole B1 are presented in Table 9. The accelerator field quality is provided within a required elliptical area with 50 mm horizontal and 110 mm vertical size (blue area in Fig. 4). It was achieved by an appropriate combination of relatively large values of low order geometrical harmonics.

Table 9: Geometrical Harmonics at $R_{ref}$=40 mm ($10^{-4}$).

| Harmonic # | Value |
|---|---|
| $b_3$ | -5.875 |
| $b_5$ | -18.320 |
| $b_7$ | -17.105 |
| $b_9$ | -4.609 |

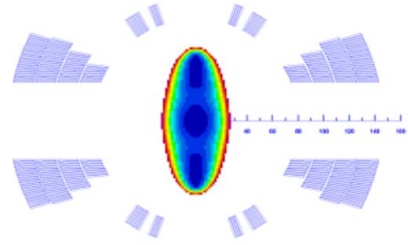

Figure 4: IR dipole cross-section.

### Radiation studies

Radiation studies for IR magnets have been started [5]. Three cases were analyzed: 1) 10-cm thick tungsten masks with $10\sigma_{x,y}$ elliptic bore placed between the IR magnets; 2) additional tungsten liners inside the quadrupoles with $10\sigma_{x,y}$ elliptic bore; 3) case 1 with the IR quadrupoles displaced horizontally by 10% of their apertures to provide ~2 T bending field. This additional field helps also to facilitate the chromaticity correction and deflect low-energy charged particles from the detector. While all the cases have advantages and limitations, it was found that a combination of all three cases may allow the peak power density in the IR magnets to be kept below their quench limits.

## CONCLUSIONS

Conceptual designs of superconducting magnets for the Storage ring of a Muon Collider with a 1.5 TeV c.o.m. energy and an average luminosity of $10^{34}$ cm$^{-2}$s$^{-1}$ were developed based on state-of-the-art $Nb_3Sn$ strands and Rutherford cables. The magnets are designed to operate at 4.5 K and provide specified operating field/field gradient with some margin and accelerator field quality in the magnet aperture.

All the magnets have quite large horizontal and vertical Lorentz force components. Both components need to be supported by adequate mechanical structures to minimize turn motion which may cause magnet quenching and field quality degradation. Handling the vertical Lorentz force component is the main challenge of the open mid-plane designs. Practical solutions to this problem, the study and optimization of magnet operating margin and other parameters will be done during short model R&D phase.